# How AI Load Swings Reduce Datacenter Power Conversion Efficiency

*H. Joshi*, A. Alenezi*, F.C. Lin*, Prasad Enjeti*, X. Wang** and X. Chen***
**Power Electronics & Power Quality Lab, **Smart power Energy & Decision Making Lab;*
http://tx.ag/TAMUCall
*Department of Electrical and Computer Engineering*
*Texas A&M University, College Station, Texas – 77843*
*Email:* enjeti@tamu.edu

**Abstract:**
AI training and inference workloads make GPU power swing repeatedly between compute and communication phases. As a result, the cascaded converters in a datacenter rack keep moving between operating points with different efficiencies. This paper uses an energy based, cycle averaged method to quantify that effect along a 240 V AC to 400 V DC to 51 V DC rack power path, using measured efficiency curves of commercially available GaN based PFC and LLC converters. For a 1 Hz, 50% duty-cycle load swinging between 5 kW and 510 W, conversion efficiency falls to 95.97%, compared with 96.82% at the equivalent constant average load, a loss of 0.88%. When 51/48V to 1V point of load conversion to the chip rail is included, it is shown that the loss grows to 1.66% (89.45% vs. 90.96%). These results show that efficiency rated at average load overstates real world performance, and that the penalty grows with each load dependent conversion stage between the grid and the GPU.

## 1. Introduction

Modern data centers increasingly rely on high density computing platforms to support Artificial Intelligence (AI) training and inference workloads. These systems require substantial DC power at a rack level for powering the AI GPUs, while the utility or facility distribution is typically provided as AC power (240/277 V) and/or 800 V DC [1]. Consequently, multiple power conversion stages are required between the facility input and computing hardware which operate at nearly at 1 V (see Fig. 1). Although each stage can be designed to achieve high efficiency under a steady operating condition, the overall rack level efficiency is determined by the combined behavior of the cascaded converters across the actual AI load profile. Fig.1 illustrates an example "Grid to Chip" [1] low voltage ac power delivery architecture for data center and highlights the conversion stages considered in this work. Although the complete path includes the facility transformer and online UPS, the present analysis begins at the 240 V AC rack input. The input is converted to a regulated 400 V DC bus through an AC-DC power factor correction stage, followed by isolated DC-DC conversion to approximately 48-51 V. The downstream processor level point of load (POL) conversion to approximately 1 V [2].

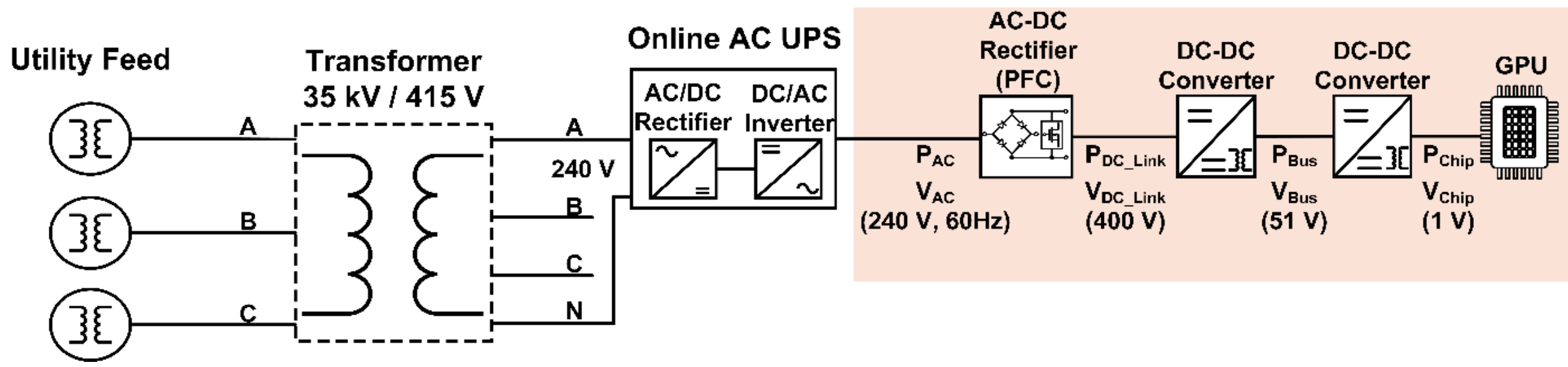

*Figure 1. An example "Grid to Chip" low voltage ac power conversion architecture employing a line frequency transformer (35 kV to 415/480 V) followed by an online UPS system with battery backup and ac-dc PFC stage, dc-dc converter and a POL powering the GPU.*

## 2. AI Rack Efficiency and Load Characterization

Fig. 2(a) shows a typical AI data center load profile between the compute and communicate phase [3]. Such variation continuously shifts the operating points of the power converters involved, causing their losses and efficiencies to vary over time [4]. To isolate the effect of these load variations on power conversion efficiency, the complex AI workload is simplified in this study as a two-level square wave load with 50% duty cycle as shown in Fig. 2(b). The high load level corresponds to the rated output power of 5 kW at 1 V, that yields 5000 A ($I_H$) load. The low load level is selected as 510 A($I_L$).

The load fluctuates at 1 Hz frequency, at a 50% duty cycle. The average load can be calculated as:

$$I_{avg} = DI_H + \ (1 - D)I_L \tag{1}$$

Where D is the fraction of each load period spent at the high load condition, and $I_H$ and $I_L$ are the high and low load currents respectively (See Fig. 2(b)). Hence, the average load current is 2755 A at 1 V. This 50% fluctuating load can be compared with the constant load current of 2755 A as both the cases would maintain the same average output power.

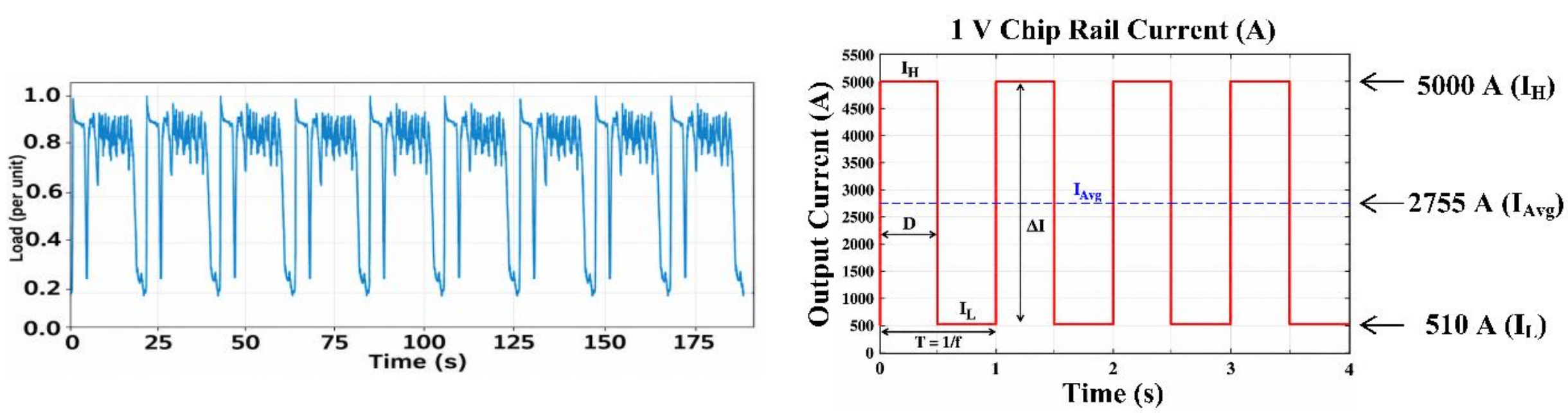

(a) (b)

*Figure 2. (a) AI Compute / Communicate load profile reported by many AI data centers [3], (b) Square wave load Profile at 50 % Duty Cycle*

### 3. Efficiency Analysis for Power Conversion Systems

To evaluate the influence of load fluctuations on rack level conversion efficiency, the efficiency characteristics of the individual power conversion stages are obtained from their respective manufacturer datasheet efficiency curves [2] [5] [6]. Fig. 3(a) and Fig. 3(b) shows the efficiency characteristics of the 240V AC to 400 V DC PFC stage and the 400 V to 51 V LLC stage, respectively, while Fig. 3(c) represents the downstream 48 V to 1 V point of load conversion [2]. All efficiency characteristics are expressed as a function of output power to enable direct comparison between the conversion stages. Alternatively, when measured efficiency characteristics are unavailable, a Vorperian based model can be used to estimate the DC-DC converter efficiency over its operating range [7].

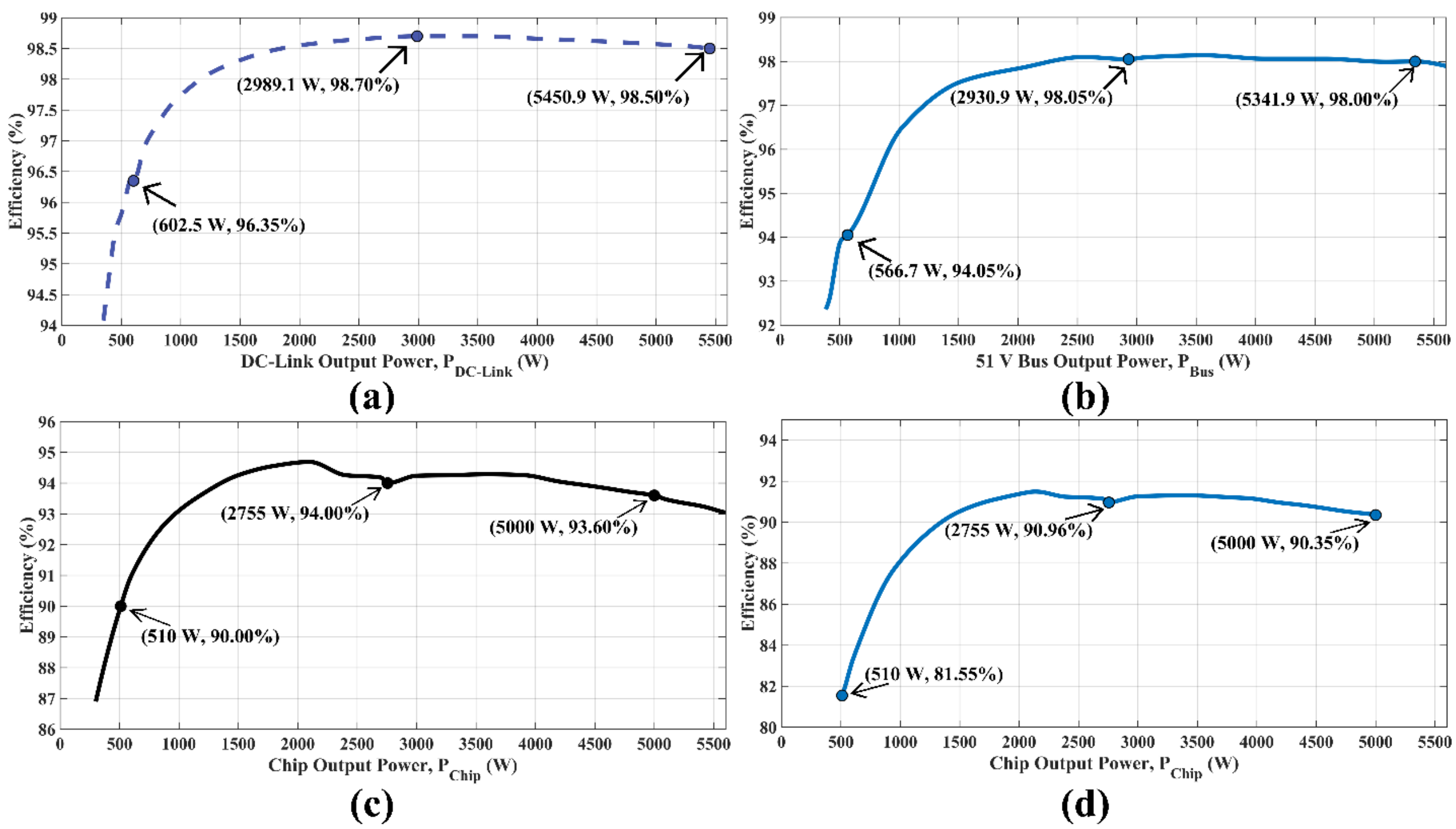

*Figure 3. Efficiency curves from the manufacturer specification sheets for the "Grid to Chip" architecture shown in Fig. 1. (a) Efficiency Curve for AC-DC Conversion stage [5], (b) Efficiency Curve for DC-DC LLC Conversion stage [6], (c) Efficiency high current Point of Load DC-DC Converter (3 in parallel) [2], (d) Overall Efficiency Curve for Combined cascaded power conversion system from 240V (AC) to 1V (DC) (multiplication of individual power conversion stages (a), (b) and (c) respectively)*

From Fig. 3 it is clear that the efficiency of various stages in the "Grid to Chip" architecture shown in Fig. 1 varies considerably as a function of the output load. At a given operating point $x$, the cascaded efficiency from 240 V AC to the 1 V Chip level is

$$\eta_{sys,x} = \eta_{POL,x} \times \eta_{DC-DC,x} \times \eta_{AC-DC,x} = \frac{P_{Chip,x}}{P_{AC,x}} \quad ; \quad x \in \{H, L, Avg\} \qquad (2)$$

Where $x$ represents the high load, low load or equivalent constant average load. The resulting steady state cascaded efficiency characteristic is shown in Fig. 3(d).

For load variation from $I_H$ to $I_L$ as shown in Fig. 2(b), the cycle efficiency for fluctuating load can be computed based on energy based, cycle average method as follows:

$$\eta_{cycle} = \frac{D * P_H + (1 - D) * P_L}{\frac{D * P_H}{\eta_H} + \frac{(1 - D) * P_L}{\eta_L}} \tag{3}$$

In Eqn (3) $\eta_H$ and $\eta_L$ represent efficiency at $I_H$ and $I_L$ load levels as depicted in Fig. 2(b). For D = 50%, $P_H$ = 5000 W, $P_L$ = 510 W, and $P_{Avg}$ = 2755 W. The corresponding power and efficiency values at these three operating points obtained from Fig. 3 are summarized in Table 1.

*TABLE 1 Power Characteristics of each conversion stage at different load levels (see Fig. 1)*

| $P_{chip}$(W) | $P_{Bus}$(W) | $P_{DC_Link}$(W) | $P_{AC}$(W) | $\eta_{POL}$(%) | $\eta_{DC-DC}$(%) | $\eta_{AC-DC}$(%) | $\eta_{Sys}$(%) |
|---|---|---|---|---|---|---|---|
| 5000 ($P_H$) | 5341.9 | 5450.9 | 5533.9 | 93.6 | 98 | 98.5 | 90.35 ($\eta_H$) |
| 510 ($P_L$) | 566.7 | 602.5 | 625.3 | 90 | 94.05 | 96.35 | 81.55 ($\eta_L$) |
| 2755 | 2930.9 | 2989.1 | 3028.5 | 94 | 98.05 | 98.7 | 90.96 |

The efficiency of the pulsating load $I_H$ to $I_L$ as shown in Fig. 2(b) can be computed from Eqn (3) and Table 1 using the energy based, cycle averaged method as follows:

$$\eta_{cycle} = \left( \frac{0.5 * 5000 + (1 - 0.5) * 510}{\frac{0.5 * 5000}{0.9035} + \frac{(1 - 0.5) * 510}{0.8155}} \right) * 100 = 89.45\ \% \tag{4}$$

Further form Table 1 (Fig. 3(d)) the average load of 2755 W has an efficiency of 90.96 %. It is clear from Eqn (4) that the AI load fluctuations decrease the overall efficiency for the same average load by 1.66 %.

## 4. Conclusion

This article presents an energy based, cycle averaged methodology to compute the efficiency of power conditioning stages for fluctuating AI training and inference workloads. It is shown that the overall efficiency is reduced by 1.66%, for example load fluctuating from 5kW to 0.5kW. The loss in efficiency contributes to additional heating and cooling burden. In the example discussed the efficiency of grid side transformer and UPS components have not been considered. Including these in calculations will further contribute to additional losses due to fluctuating AI loads.

**The takeaway:** In an hyperscale AI datacenter, average load efficiency is an optimistic fiction, because every swing between compute and communicate costs watts the datasheet never shows, it is not the average load that sets efficiency, but how the load moves. Furthermore, at gigawatt scale, a 1.66% loss in efficiency amounts to a small power plant's worth of lost energy.

## References

[1] Nvidia, "NVIDIA 800 VDC Architecture," [Online]. Available: https://www.nvidia.com/en-us/data-center/technologies/800-vdc-architecture/. [Accessed 14 August 2026].

[2] José A Cobos, "High-Current Power Delivery for AI Accelerators: The Case for Direct Power Conversion," in *Energy Conversion Congress & Expo (ECCE) Europe*, Valencia, Spain, 2026.

[3] E. Choukse, et al., "Power Stabilization for AI Training Datacenters," *arXiv:2508.14318v1 [cs.AR],* 2025.

[4] A. Pratt, P. Kumar and T. V. Aldridge, "Evaluation of 400V DC distribution in telco and data centers to improve energy efficiency," in *INTELEC 07 - 29th International Telecommunications Energy Conference*, Rome, Italy, 2007.

[5] Efficient Power Conversion, "EPC91107KIT: 5 kW 4-Level Totem-Pole PFC with EPC2304 GaN FETs," 2025. [Online]. Available: https://epc-co.com/epc/portals/0/epc/documents/guides/EPC91107_qsg.pdf. [Accessed 07 October 2026].

[6] Efficient Power Conversion, "EPC91110KIT: 5.5 kW, 8:1 fixed ratio, 400 V to 50 V Isolated Converter with EPC2305 GaN FETs," 2025. [Online]. Available: https://epc-co.com/epc/portals/0/epc/documents/guides/EPC91110_qsg.pdf. [Accessed 07 October 2026].

[7] V. Vorpérian, "Simple Efficiency Formula for Regulated DC-to-DC Converters," *IEEE Transactions on Aerospace and Electronic Systems,* vol. 46, no. 4, pp. 2123-2131, 2010.